\documentclass[fleqn,twoside]{article}
\usepackage{espcrc2}

\usepackage{graphicx}
\usepackage[figuresright]{rotating}

\newcommand{\AmS}{{\protect\the\textfont2
  A\kern-.1667em\lower.5ex\hbox{M}\kern-.125emS}}
\newcommand{\gsim}{\raisebox{-0.07cm  }
{$\, \stackrel{>}{{\scriptstyle\sim}}\, $}}
\newcommand{\ep}{\varepsilon}
\newcommand{\N}{\nonumber}
\newcommand{\D}{\displaystyle}

\newcommand{\SigmaP}{{\sf Sigma}}
\newcommand{\eph}{\frac{\ep}{2}}

\title{\vspace*{-4mm}
{\tiny DESY 10--093 \hfill SFB/CPP--10--58 \hfill  IFIC/10--22 \hfill  TTK--10--38}\\
Heavy Flavor DIS Wilson coefficients in the asymptotic regime}

\author{
 J. Ablinger\address[LINZ]{Research Institute for Symbolic Computation (RISC),\\
                          Johannes Kepler University, Altenbergstra\ss e 69, 
                          A--4040, Linz, Austria},
 I. Bierenbaum\address{Instituto de F\'{i}sica Corpuscular, CSIC-Universitat 
                       de Val\`{e}ncia, \\ 
                       Apartado de Correos 22085, E-46071 Valencia, Spain},
 J. Bl\"umlein\address[ZEUTHEN]{Deutsches Elektronen--Synchrotron, DESY,
                               Platanenallee 6, D--15738, Germany},
 A. Hasselhuhn\addressmark[ZEUTHEN],
 S. Klein\address{Institut f\"ur Theoretische Teilchenphysik und Kosmologie, \\
                 RWTH Aachen University, D--52056 Aachen, Germany}
                 \thanks{Speaker},
 C. Schneider\addressmark[LINZ],
 F. Wi\ss brock\addressmark[ZEUTHEN]
  }

\begin{document}
\thispagestyle{empty}

\begin{abstract}
 \noindent
 We report on results for the heavy flavor contributions to $F_2(x,Q^2)$ in the limit 
 $Q^2\gg m^2$ at {\sf NNLO}. By calculating the massive $3$--loop operator matrix elements, 
 we account for all but the power suppressed terms in $m^2/Q^2$. Recently, the calculation 
 of fixed Mellin moments of all $3$--loop massive operator matrix elements has been finished. 
 We present new all--$N$ results for the $O(n_f)$--terms, thereby confirming the corresponding 
 parts of the $3$--loop anomalous dimensions. Additionally, we report on first genuine $3$--loop 
 results of the ladder--type diagrams for general values of the Mellin variable $N$.
\vspace{1pc}
\end{abstract}

\maketitle
%
%
%

\section{Introduction}
\noindent
Deep-inelastic scattering (DIS) in the region of large enough values of 
the gauge boson virtuality $Q^2 = -q^2$, allows to measure
the leading twist parton densities of the nucleon, the QCD-scale
$\Lambda_{\rm QCD}$, and, related to this, the strong coupling constant $a_s(Q^2) =
\alpha_s(Q^2)/(4\pi)$, to high precision, \cite{ALPS}. Especially in the region of
smaller values of Bjorken--$x$, the DIS--structure functions $F_{2,L}(x,Q^2)$
contain large $c\overline{c}$--contributions of up to 20-40~\%, therefore deserving 
further investigation. Our goal is the calculation of the 
completely inclusive heavy flavor Wilson coefficients which constitute the 
perturbative contributions to the structure functions and are denoted 
by ${\sf H}_{j,(2,L)}(x,Q^2/\mu^2,m^2/\mu^2)$. Here, 
we include effects of one species of heavy flavors in the final state and/or
its virtual contributions.
At present, the heavy flavor Wilson coefficients
are known exactly to ${\sf NLO}$ by a semi--analytically result in $x$--space \cite{HEAV1}.
Due to the size of the heavy flavor corrections, it is necessary to extend 
the description of these contributions to $O(a_s^3)$, and thus to the
same level which has been reached for the massless Wilson 
coefficients \cite{Vermaseren:2005qc}.

\noindent
A calculation of
these quantities in the whole kinematic range at ${\sf NNLO}$ seems to be out
of reach at present. However, in the limit of large virtualities,  $Q^2\:
\gsim \:10\:m_c^2$ in the case of $F_2^{c\bar{c}}(x,Q^2)$, one observes that
$F_{2,L}^{c\bar{c}}(x,Q^2)$ are very well described by their asymptotic
expressions \cite{Buza:1995ie} disregarding power corrections in $m^2/Q^2$. 
In this kinematic range, the heavy flavor Wilson coefficients have been 
obtained analytically for $F^{c\bar{c}}_2(x,Q^2)$ to 2--loop order
in \cite{Buza:1995ie,BBK1} and for $F^{c\bar{c}}_L(x,Q^2)$ to 3--loop order in
\cite{Blumlein:2006mh}. The asymptotic expressions are obtained by 
a factorization of the heavy quark Wilson coefficients into a Mellin 
convolution of massive operator matrix elements (OMEs) $A_{jk}$ and the 
massless Wilson coefficients $C_{j,i}$. In case of only one heavy flavor species 
this factorization reads in Mellin--space, \cite{Buza:1995ie,Buza:1996wv},
 \begin{eqnarray}
    && {\sf H}_{j,(2,L)}(N)
        = A_{ij}(N) \cdot C_{i,(2,L)}(N)~,\N\\
       &&\hspace{45mm}i,j=q,g~. \label{CallFAC}
 \end{eqnarray} 
Here, we indicated the dependence on Mellin--$N$ and suppressed
all further variables, cf. \cite{Buza:1995ie,Bierenbaum:2009mv} for details.
Eq. (\ref{CallFAC}) allows to calculate the heavy fla- \newline vor Wilson coefficients
in the limit $Q^2\gg m^2$ up to $O(a_s^3)$ by combining the results obtained
in Ref. \cite{Vermaseren:2005qc} for the light flavor Wilson coefficients with
the $3$--loop massive OMEs. 

\noindent 
Another physical motivation for this calculation is that the
massive OMEs also serve as transition functions if one wants to define parton 
densities for massive quarks in the framework of a variable flavor number 
scheme, \cite{Buza:1996wv}. This is of particular interest for heavy quark
induced processes at the LHC, such as $c\overline{s}\rightarrow W^+$ at large
enough scales $Q^2$. Another important point is that in the course of our
calculation, we obtain the $O(n_f)$--parts of the $3$--loop anomalous 
dimensions, \cite{ANDIMNSS}, which are thus confirmed for the first time 
in an independent calculation. Finally we are also interested in the 
mathematical structures of the Feynman-parameter integrals emerging in our calculation, leading
to new insight, cf. \cite{CarstenLL}.

\noindent
In the following, we briefly 
describe the calculation of the fixed moments 
of the massive OMEs and then present our recently obtained all--$N$ results.
For more details on the challenges we encountered on the mathematical side, see
~\cite{CarstenLL}.
%
%
%
\section{Massive OMEs and Fixed Moments}
\noindent 
We are interested in the flavor--decomposed twist--2 massive OMEs
\begin{eqnarray}
  &&\hspace{-6mm}  A_{ki}^{\sf S,NS}\Bigl(N,\frac{m^2}{\mu^2}\Bigr)
  = \langle i|O_k^{\sf S, NS}|i\rangle_H
\N\\ && \hspace{12mm}
  = \delta_{ki} + \sum_{l=1}^{\infty} a_s^l
     A_{ki}^{{\sf S, NS},(l)}\Bigl(N,\frac{m^2}{\mu^2}\Bigr)~.
       \label{OMEs}
\end{eqnarray}
The external on--shell particles are denoted by $i=q,g$ and $O_k$
stands for the quarkonic ($k=q$) or gluonic ($k=g$) composite operator emerging
in the light--cone expansion. The subscript $H$ indicates that we require the
presence of heavy quarks of one type with mass $m$. $\mu^2$ denotes both the 
factorization and renormalization scales. The composite operators give 
rise to additional Feynman--rules depending on Mellin--$N$ which can 
be found in Ref. \cite{Bierenbaum:2009mv}. 

\noindent
In case of the gluon operator, the contributing terms are denoted by
$A_{gq,Q}$ and $A_{gg,Q}$. For the quark operator, one distinguishes whether 
the current couples to a heavy or light quark. In the 
non--singlet (${\sf NS}$)--case, the
operator, by definition, couples to the light quark. Thus there is only one 
term, $A_{qq,Q}^{\sf NS}$. In the singlet (${\sf S}$) and pure--singlet 
(${\sf PS}$)--case, two OMEs can
be distinguished, $\D{\{A_{qq,Q}^{\sf PS},~A_{qg,Q}^{\sf S}\}}$
and $\D{\{A_{Qq}^{\sf PS},~A_{Qg}^{\sf S}\}}$, where, in the former case, 
the operator couples to a light quark and in the latter case to a heavy 
quark. We also consider the transversity operator, giving rise to the 
OME $\Delta_T A^{\sf NS}_{qq,Q}$~.

\noindent 
Up to and including 2--loop order, all massive OMEs have been calculated in
Refs. \cite{Buza:1995ie,Buza:1996wv} and confirmed independently
in \cite{BBK1,Bierenbaum:2009zt}. The transversity terms have been 
obtained in Ref. \cite{Blumlein:2009rg}. Additionally, 
all 2--loop $O(\ep)$--terms in $D=4+\ep$ dimensions
have been calculated in 
\cite{Bierenbaum:2009zt,Blumlein:2009rg,Bierenbaum:2008yu}. 
These terms are needed for the renormalization of the massive OMEs at 3--loops. 
Let us briefly review the most important steps of 
renormalization, \cite{Buza:1995ie,Bierenbaum:2009mv}.
We apply the $\overline{\rm{MS}}$--scheme, except for mass renormalization, 
which is performed in the on--shell scheme \cite{MASS2}.  After mass and 
coupling constant renormalization, the
remaining singularities are of the
ultraviolet and collinear type. The former are renormalized via the operator
$Z$--factors, whereas the latter are removed via mass factorization 
and absorbed into the parton densities. Note that in the last two steps 
the anomalous dimensions of the twist--2 operators emerge. Thus 
at ${\sf NNLO}$ the fermionic parts of the 3-loop anomalous dimensions 
calculated in Refs. \cite{ANDIMNSS,Gracey:2003yrxGracey:2003mrxGracey:2006zrxGracey:2006ah}, cf. also \cite{Gracey:1993nn}, appear.
The general structure of the un-renormalized and renormalized 
OMEs at $3$--loops is 
\begin{eqnarray}
 \hat{\hspace*{-1mm}\hat{A}}_{ij}^{(3)}\Bigl(N,\frac{m^2}{\mu^2}\Bigr)
   \!\!\!\!&=&\!\!\!\!
   \Bigl(\frac{m^2}{\mu^2}\Bigr)^{\frac{3\ep}{2}}
    \sum_{i=1}^3\frac{a_{ij}^{(3),-i}}{\ep^i}+a_{ij}^{(3)}
      ~, \label{Ahatij3}  \\
 A_{ij}^{(3)}\Bigl(N,\frac{m^2}{\mu^2}\Bigr)\!\!\!\!&=&\!\!\!\!
    \sum_{i=0}^3 
        a_{ij}^{(3),i}\ln^i\Bigl(\frac{m^2}{\mu^2}\Bigr)~. 
       \label{Aij3}
\end{eqnarray}
The pole terms in $\ep$ of Eq. (\ref{Ahatij3}) and the logarithmic 
terms in $m^2/\mu^2$ in Eq. (\ref{Aij3}), respectively, 
are completely determined by renormalization and can be expressed in terms
of the anomalous dimensions up to $3$ loops, 
the expansion coefficients of the QCD $\beta$--function up to $2$ loops and 
the $1$-- and $2$--loop contributions to the massive OMEs. For the exact 
formulae, we refer to \cite{Bierenbaum:2009mv}. Hence these
terms are already known for general values of $N$ and can be used 
for first phenomenological analyses, \cite{BBKPREP}. 
This is not the case for the constant term, which contains the 
genuine $3$--loop contributions $a_{ij}^{(3)}$.

\noindent
The massive OMEs at $O(a_s^3)$ are given by $3$--loop self--energy type
diagrams, which contain a local operator insertion. The external massless
particles are on--shell. The heavy quark mass sets the scale and the spin of
the local operator is given by the Mellin--variable $N$. 
Using the programs {\sf QGRAF}, \cite{Nogueira:1991ex}, {\sf color}, 
\cite{vanRitbergen:1998pn}, {\sf MATAD}, \cite{Steinhauser:2000ry}, 
and the computer algebra system {\sf FORM}, \cite{Vermaseren:2000nd}, we calculated fixed moments 
of all OMEs. For the terms $A_{Qg}^{(3)}, A_{qg,Q}^{(3)}$ and
$A_{gg,Q}^{(3)}$ the even moments $N = 2$ to 10, for $A_{Qq}^{(3), \rm PS}$ to 
$N = 12$, and for 
$A_{qq,Q}^{(3), \rm NS}$, $A_{qq,Q}^{(3),\rm PS}$, $A_{gq,Q}^{(3)}$ to $N=14$ 
were computed.
For the flavor non-singlet terms, we calculated as well the odd moments $N=1$ 
to $13$, corresponding to the light flavor $(-)$-combinations, 
\cite{Bierenbaum:2009mv}. We 
also calculated the transversity OME $\Delta_T A^{(3),\rm NS}_{qq,Q}$ for 
the even/odd moments $N=1..13$, \cite{Blumlein:2009rg}.
All our results agree with the predictions obtained from renormalization, 
providing us with a strong check on our calculation.
\noindent
For all moments we calculated, we agree with the known results in 
the literature, especially with the $T_f$--parts of the $3$--loop anomalous
dimensions, \cite{ANDIMNSS,Gracey:2003yrxGracey:2003mrxGracey:2006zrxGracey:2006ah,Gracey:1993nn}.
Our complete results are given in Refs. \cite{Bierenbaum:2009mv,Blumlein:2009rg}
and will be used for first phenomenological parameterizations, \cite{BBKPREP}. 

%
%
%
\section{The $n_f$--contributions for all--N}
\noindent 
As a first step towards the calculation of 
the complete $3$--loop massive OMEs for general values of $N$,
we consider the $n_f$--contributions only.
In the following we describe the computation of these terms and present our
results for the constant term $a_{Qg}^{(3)}$
and the $3$--loop anomalous dimension $\gamma_{qg}^{(2)}$ which enters the 
single pole term in Eq. (\ref{Ahatij3}). The remaining OMEs will be given
in \cite{ABKSW1}, cf. also \cite{FABDIPL}.

\noindent
In Refs. \cite{BBK1,Bierenbaum:2009zt,Bierenbaum:2008yu},
the $2$--loop massive OMEs were calculated
using Feynman--parameters in order to arrive at a representation of 
the momentum integrals in terms of finite and infinite 
sums which depend on the Mellin--variable $N$. The summand is a
hypergeometric expression in terms of $\Gamma$--functions, 
which could safely be expanded in $\ep$, before
performing the summation. Thus we obtained up to twofold sums, 
one of which derived from a hypergeometric function 
$\empty_3F_2$ extending to infinity and the other usually being a 
finite sum. These sums could then be 
calculated applying algebraic or analytic summation techniques and 
for the more complicated sums the summation package \SigmaP, 
\cite{CarstenLL,Bierenbaum:2008yu,sigma}. 

\begin{figure}[htb]
      \begin{center}
       \includegraphics[angle=0, width=6.5cm]{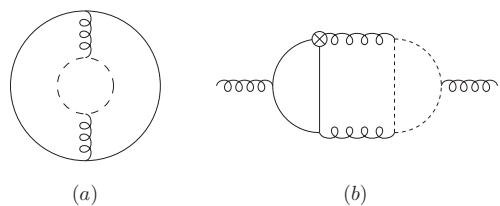}
      \end{center}

\vspace*{-8mm}
       \caption{\sf Examples for $3$--loop diagrams $\propto~n_f$. Solid 
                 lines: heavy quarks, dashed lines: light quarks, curly lines:
                 gluons. $\otimes$: operator insertion.}
       \label{3LOOPnf}
     \end{figure}

\noindent
For the $3$--loop $n_f$--terms, we followed in principle the same approach. 
In this case an additional light quark loop is present
along with the heavy quark loop, compared to the $2$--loop case. 
There is one basic topology only, which is shown in 
Figure~\ref{3LOOPnf}(a). From this all contributing diagrams can be 
derived by 
attaching outer light partons or ghosts and inserting the quarkonic operator in
all possible ways. Since the external particles are on--shell, for the lowest moment
all diagrams can be reduced to this massive tadpole. By recursion, this 
holds for higher moments as well. As an example, in Figure \ref{3LOOPnf}, (b), 
one of the more complicated diagrams contributing to the 
$n_f$--part of $A_{Qg}^{(3)}$ is shown, cf. \cite{CarstenLL}.

\vspace*{-5mm}

\vspace*{-5mm}
\noindent
The calculation of the massive tadpole shown in Figure \ref{3LOOPnf}, (a), is 
straightforward. After integrating out the 
light quark loop, one is left with a $2$--loop massive tadpole, for which 
the result is known analytically for arbitrary exponents of the propagators. 
However, as in the $2$--loop case, the calculation is significantly more 
complicated once the operator insertion is present. 
Nevertheless, all momentum integrals could be
re--written in terms of up to two fourfold sums.
These sums differed in their structure from those
encountered in the $2$--loop case, making the calculation significantly more 
complicated. Here, we refer to \cite{CarstenLL} for more details on 
how these sums were calculated using the {\sf MATHEMATICA}--based 
program \SigmaP, \cite{sigma}.

\noindent 
Let us now present some of our results. For the $n_f^2$--term
of the $3$--loop anomalous dimension $\gamma_{qg}^{(2)}$ 
we obtain\footnote{Note that the term $\propto~n_f^2$ in \cite{ANDIMNSS} corresponds for the massive 
OMEs to the term $\propto~n_f$.}
%
%
%
\begin{eqnarray} 
 \gamma_{qg}^{(2)}&=&
    n_f^2T_f^2C_A\Biggl\{
        \frac{32(N^2+N+2)}{9N(N+1)(N+2)}\Bigl[
            12S_{2,1}
\N\\ && \hspace{-12mm}
           +12S_{-3}
           +2S_3
           -3S_2S_1
           +S_1^3
                                     \Bigr]
\N\\ && \hspace{-12mm}
       -\frac{128(5N^2+8N+10)}{9N(N+1)(N+2)}S_{-2}
\N\\ && \hspace{-12mm}
       -\frac{64P_1S_2+64P_2S_1^2}{9N(N+1)^2(N+2)^2}
\N\\ && \hspace{-12mm}
       +\frac{64P_3S_1}{27N(N+1)^3(N+2)^3}
\N\\ && \hspace{-12mm}
       +\frac{16P_4}{27(N-1)N^4(N+1)^4(N+2)^4}
                   \Biggr\}
\N\\ && \hspace{-12mm}
     +       n_f^2T_f^2C_F\Biggl\{
        \frac{32(N^2+N+2)}{9N(N+1)(N+2)}\Bigl[
            10S_3
\N\\ && \hspace{-12mm}
           -3S_1S_2
           -S_1^3
                                     \Bigr]
       +\frac{32(5N^2+3N+2)}{3N^2(N+1)(N+2)}S_2
\N\\ && \hspace{-12mm}
       +\frac{32(10N^3+13N^2+29N+6)}{9N^2(N+1)(N+2)}S_1^2
\N\\ && \hspace{-12mm}
       -\frac{32P_5S_1}{27N^2(N+1)^2(N+2)}
\N\\ && \hspace{-12mm}
       +\frac{4P_6}{27(N-1)N^5(N+1)^5(N+2)^4}
                          \Biggr\}~, \label{gammaqg}
\end{eqnarray}
with the polynomials
\begin{eqnarray}
%
P_1&=&5N^4+26N^3+47N^2+43N+20~, \\
P_2&=&5N^4+20N^3+41N^2+49N+20~, \\
P_3&=&19N^6+124N^5+492N^4+1153N^3
\N\\ && \hspace{-12mm}
        +1362N^2+712N+152~, \\
P_4&=&1152+7296N+19904N^2
\N\\ && \hspace{-12mm}
      +30864N^3+25896N^4+6800N^5
\N\\ && \hspace{-12mm}
      -10364N^6-8899N^7+3557N^8+8534N^9
\N\\ && \hspace{-12mm}
      +5194N^{10}+1485N^{11}+165N^{12}~, \\
P_5&=&47N^4+145N^3+426N^2+412N
\N\\ && \hspace{-12mm}
       +120~, \\
P_6&=& 99N^{14}+990N^{13}+4925N^{12}
\N\\ && \hspace{-12mm}
       +17916N^{11}+46649N^{10}+72446N^9
\N\\ && \hspace{-12mm}
        +32283N^8-95592N^7-267524N^6
\N\\ && \hspace{-12mm}
        -479472N^5-586928N^4-455168N^3
\N\\ && \hspace{-12mm}
        -269760N^2-122112N-27648~.
\end{eqnarray}
The above and the following expressions are given in terms of harmonic
sums, \cite{Vermaseren:1998uu,Blumlein:1998if}, 
taken at argument $N$. Note that we perform an analytic continuation to 
complex values of $N$ starting from the even moments. Eq. (\ref{gammaqg})
agrees with the corresponding results in the literature for the case of general values of $N$
and for fixed moments \cite{Bierenbaum:2009mv,ANDIMNSS,Gracey:1993nn}. 
Our new result is the constant term, which reads
%
%
\begin{eqnarray}
 a_{Qg}^{(3)}&=& 
    n_fT_f^2C_A\Biggl\{
         \frac{16(N^2+N+2)}{27N(N+1)(N+2)}\Bigl[
            -6 S_{3,1}
\N\\ && \hspace{-12mm}
            +108 S_{-2,1,1}
            -78 S_{2,1,1}
            -90 S_{-3,1}
            +72 S_{2,-2}
\N\\ && \hspace{-12mm}
            -108 S_{-2,1} S_1
            +42 S_{2,1} S_1
            -6 S_{-4}
            +90 S_{-3} S_1
\N\\ && \hspace{-12mm}
            +118 S_3 S_1
            +120 S_4
            +18 S_{-2} S_2
            +54 S_{-2} S_1^2
\N\\ && \hspace{-12mm}
            +33 S_2 S_1^2
            +15 S_2^2
            +2 S_1^4
            +18S_{-2}\zeta_2
            +9S_2\zeta_2
\N\\ && \hspace{-12mm}
            +9S_1^2\zeta_2
            -42 S_1\zeta_3
                                    \Bigr]
        -\frac{64Q_{1}S_{2,1}+16Q_{2}S_2S_1}{27N(N+1)^2(N+2)^2}
\N\\ && \hspace{-12mm}
        +\frac{32Q_{3}(6S_{-2,1}-5S_{-3}-6S_{-2}S_1)}{27N(N+1)^2(N+2)^2}
\N\\ && \hspace{-12mm}
        -\frac{16Q_{4}S_1^3+144Q_{5}S_1\zeta_2}{81N(N+1)^2(N+2)^2}
\N\\ && \hspace{-12mm}
        +\frac{32Q_{6}S_{-2}+8Q_{7}S_1^2}{81N(N+1)^3(N+2)^3}
\N\\ && \hspace{-12mm}
        -\frac{32Q_{8}S_3-4032Q_{9}\zeta_3}{81(N-1)N^2(N+1)^2(N+2)^2}
\N\\ && \hspace{-12mm}
        +\frac{Q_{10}S_2-9Q_{11}\zeta_2}{81(N-1)N^3(N+1)^3(N+2)^3}
\N\\ && \hspace{-12mm}
        +\frac{N^3(N+1)(N+2)Q_{12}S_1-Q_{13}}{243(N-1)N^5(N+1)^5(N+2)^5}
       \Biggr\}
\N\\ && \hspace{-12mm}
   +n_fT_f^2C_F\Biggl\{
         \frac{16(N^2+N+2)}{27N(N+1)(N+2)}\Bigl[
             144S_{2,1,1}
\N\\ && \hspace{-12mm}
            -72S_{3,1} 
            -72S_{2,1}S_1
            +48S_4
            -16S_3S_1
            -24S_2^2
\N\\ && \hspace{-12mm}
            -12S_2S_1^2
            -2S_1^4
            -9 S_1^2\zeta_2
            +42S_1\zeta_3
                                    \Bigr]
\N\\ && \hspace{-12mm}
        +32\frac{10N^3+49N^2+83N+24}{81N^2(N+1)(N+2)}\Bigl[
             3S_2S_1
             +S_1^3
                                    \Bigr]
\N\\ && \hspace{-12mm}
        -\frac{128(N^2-3N-2)}{3N^2(N+1)(N+2)}S_{2,1}
\N\\ && \hspace{-12mm}
        -\frac{96Q_{14}(N+1)S_1^2+16Q_{15}S_1}{243N^2(N+1)^3(N+2)}
\N\\ && \hspace{-12mm}
        +\frac{16(5N^3+11N^2+28N+12)S_1\zeta_2}{9N^2(N+1)(N+2)}
\N\\ && \hspace{-12mm}
        -\frac{Q_{16}S_3+9Q_{17}\zeta_3}{81(N-1)N^3(N+1)^3(N+2)^2}
\N\\ && \hspace{-12mm}
        +\frac{Q_{18}S_2+3Q_{19}\zeta_2}{27(N-1)N^4(N+1)^4(N+2)^3}
\N\\ && \hspace{-12mm}
        +\frac{Q_{20}}{243(N-1)N^6(N+1)^6(N+2)^5} 
        \Biggr\}~.\label{aQg3}
\end{eqnarray}
The terms $Q_1\ldots Q_{20}$ are rather lengthy irreducible polynomials 
in $N$ which will be given in Ref.~\cite{ABKSW1}.
Eq.~(\ref{aQg3}) agrees for fixed moments with the results 
of Ref. \cite{Bierenbaum:2009mv}. 
The term $A_{Qg}^{(3)}$ forms the most complex contribution to the 
$O(n_f)$ terms of the massive OMEs, along with the term
$A_{qg,Q}^{(3)}$. The latter as well as the $O(n_f)$--terms of 
$A_{Qq}^{{\sf PS}, (3)}$, $A_{qq,Q}^{{\sf PS}, (3)}$, 
$A_{qq,Q}^{{\sf NS}, (3)}$ and $\Delta_T A_{qq,Q}^{{\sf NS}, (3)}$ have been 
obtained by us as well and will be presented in \cite{ABKSW1} in detail.
Furthermore, the corresponding contributions to the remaining 3-loop anomalous
dimensions in the vector-- and transversity case are obtained. In this way, 
an independent recalculation of these quantities given in 
\cite{ANDIMNSS,Gracey:2003yrxGracey:2003mrxGracey:2006zrxGracey:2006ah,Gracey:1993nn} before, was performed using very different methods.

\noindent Finally, it is interesting to consider the 
small-- and large--$N$ limits. We obtain for the $n_f$--terms of the 
renormalized OME $A_{Qg}^{(3)}$
\begin{eqnarray}
 &&\hspace{-6mm}\lim_{N \rightarrow 1} (N-1)A_{Qg}^{(3)}=
  T_f^2n_f\Biggl\{ 
       \Biggl(
           -\frac{64}{27}C_A
           -\frac{128}{27}C_F
       \Biggr)
\N\\ &&\hspace{-6mm}
       \times \ln^3 \Bigl  (\frac{m^2}{\mu^2}\Bigr)
      +\Biggl(
            \frac{416}{27}C_A
           -\frac{832}{27}C_F
       \Biggr)\ln^2 \Bigl  (\frac{m^2}{\mu^2}\Bigr)
\N\\ &&\hspace{-6mm}
      +\Biggl(
           -\frac{32}{27}C_A
           -\frac{8512}{81}C_F
       \Biggr)\ln \Bigl  (\frac{m^2}{\mu^2}\Bigr)
      +C_A\Bigl(\frac{13088}{729}
\N\\ &&\hspace{-6mm}
      +\frac{512}{27}\zeta_3\Bigr)
      +C_F\Bigl(-{\frac{122432}{729}}+{\frac{1024}{27}}\zeta_3\Bigr)
          \Biggr\}~, 
\\ &&\hspace{-6mm}
      \lim_{N \rightarrow \infty}A_{Qg}^{(3)}=
  T_f^2n_f\frac{C_F-C_A}{N}\Biggl\{ 
           -\frac{20}{27} \hat{\ln}^4(N)
\nonumber\\ &&\hspace{-6mm}
           +\Biggl(
               -\frac{32}{9}\ln \Bigl(\frac{m^2}{\mu^2}\Bigr)
               +\frac{320}{81}
            \Biggr) \hat{\ln}^3(N)
                          \Biggr\}
\nonumber\\&& \hspace{-6mm}
     +O\Bigl(\frac{\hat{\ln}^2(N)}{N}\Bigr)~, 
\\&& \hspace{-6mm}
      \hat{\ln}(N)\equiv \ln(N)+\gamma_E~, 
\end{eqnarray}
with $\gamma_E$ being the Euler--Mascheroni constant.
\noindent
The $n_f $--contributions at $O(a_s^3)$ to the massive OMEs
contain nested harmonic sums up to weight {\sf w = 4}. 
This also applies to all individual Feynman diagrams, which we calculated in 
Feynman--gauge. After reducing the contributing harmonic sums to their common 
basis using algebraic relations \cite{Blumlein:2003gb}
between them,
the following harmonic sums contribute
\begin{eqnarray}
\label{eq:HSUM1}
& & S_1,~\quad S_2,~S_{-2},~\quad S_3,~S_{-3}, \N\\
& & S_{2,1},~S_{-2,1},~\quad S_4,~S_{-4}, \N\\ 
& & S_{3,1},~S_{-3,1},~S_{-2,2},~\quad S_{2,1,1},~S_{-2,1,1}~.
\end{eqnarray}
Note that this set of harmonic sums does not contain the index $\{-1\}$.
Due to the structural relations 
\cite{Blumlein:2009ta,ABS2010}, 
differentiation and 
argument-duplication, cf. \cite{Blumlein:1998if}, the 
set (\ref{eq:HSUM1}) can be reduced even further.
$S_1$ represents the class of all single harmonic sums.
Hence only the {\it six} basic harmonic sums
\begin{eqnarray}
  S_1~,S_{2,1},~S_{-2,1}~,S_{-3,1},~S_{2,1,1},~S_{-2,1,1}~.\label{eq:HSUM3}
\end{eqnarray}
are needed to represent the 3-loop results for the 
$n_f $--contributions to the OMEs. 

\noindent
In intermediate steps, we also observed so-called generalized 
harmonic sums \cite{Moch:2001zr,ABS2010}. They obey the following recursive 
definition \cite{Moch:2001zr}~:
\begin{eqnarray} 
 &&\widetilde{S}_{m_1,...}(x_1,...; N)=
 \N\\ && 
         \sum_{i_1=2}^N \frac{x_1^{i_1}}{i_1^{m_1}} 
         \sum_{i_2=1}^{i_1-1} \frac{x_2^{i_2}}{i_2^{m_2}}
         \widetilde{S}_{m_3,...}(x_3,...; i_2) 
 \N\\ & & 
         +\widetilde{S}_{m_1+m_2,m_3,...}( x_1 \cdot x_2,x_3,...; N)~. 
         \label{eq:HSUM4} 
\end{eqnarray} 
In the present calculation the values of $x_i$ extend to 
$\{-1/2, 1/2, -2, 2\}$. These sums occur in ladder like structures,
cf.~\cite{Vermaseren:2005qc,ABKSW1,BHKS} and the next section. They may 
also emerge, however, if contributions to $3$--loop Feynman diagrams 
containing a 2-point insertion are separated into various terms in an arbitrary
way. The weight of these sums can reach {\sf w = 5}, 
although only {\sf w = 4} sums should remain in the final results. 
Examples for these sums are~:
\begin{eqnarray} 
 &&\widetilde{S}_{1}(1/2;N),~~
   \widetilde{S}_2(-2;N),~~
   \widetilde{S}_{2,1}(-1,2;N),~~
   \N\\ &&
   \widetilde{S}_{3,1}(-2,-1/2;N),~~
   \widetilde{S}_{2,3}(-2,-1/2;N),~~
   \N\\&&
   \widetilde{S}_{1,1,1,2}(-1,1/2,2,-1;N),~~{\rm etc.}
   \label{eq:HSUM5} 
\end{eqnarray}
The algebraic and structural relations for these sums are worked out in 
Ref.~\cite{ABS2010}. Similar to the case of harmonic sums, corresponding 
basis representations are obtained. Expressing our results in terms 
of this common basis, all generalized harmonic sums canceled for each 
individual diagram. In the beginning, however, it has been unclear, whether
this was bound to happen. It is not excluded that these sums contribute
in the final results of some of the massive OMEs at 3--loop order.

%
%
%
\section{Ladder--type structures}
\begin{figure}[htb]
      \begin{center}
       \includegraphics[angle=0, width=4cm]{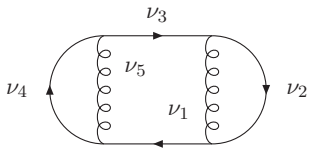}
      \end{center}

\vspace*{-8mm}
       \caption{\sf $3$--loop ladder tadpole. $\nu_i$ denote exponents of 
                the propagators.}
       \label{TOP3A}
    \end{figure}

 \noindent
 As the next step towards a complete calculation of the $3$--loop OMEs, 
 we consider diagrams deriving from the ladder--type topology shown 
 in Figure \ref{TOP3A}, cf. \cite{BHKS}. 
 Contrary to the diagram shown in Figure~\ref{3LOOPnf}(a), it can not trivially be reduced to a 
$2$--loop 
 diagram. 

\vspace*{-5mm}
\begin{figure}[htb]
      \begin{center}
       \includegraphics[angle=0, width=4.5cm]{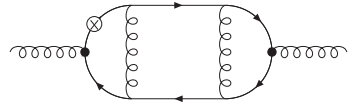}
      \end{center}

\vspace*{-8mm}
       \caption{\sf Example $3$--loop ladder diagram.}
       \label{TOP3AEX}
    \end{figure}

 However, one finds that in the scalar case it can be represented in terms of an
 Appell--function of the first kind, $F_1$,
 \begin{eqnarray}
     &&\hspace{-6mm}F_1\Bigl[a;b,b';c;x_3,x_4\Bigr] 
\N\\ &&\hspace{-6mm}
        =\sum_{m,n=0}^{\infty}
                     \frac{(a)_{m+n}(b)_n(b')_m}
                       {(1)_m(1)_n(c)_{m+n}}
                     x_3^nx_4^m~ 
      =\int_0^1dx_1
\N\\ &&\hspace{-4mm}
     \times \int_0^{1-x_1}dx_2 \frac{
      x_1^{b-1}x_2^{b'-1}(1-x_1-x_2)^{c-b-b'-1}}{(1-x_1x_3-x_2x_4)^{a}}~,\N\\
      \label{F1def}
 \end{eqnarray}
 for arbitrary exponents of the propagators. This leads to a double 
 infinite sum ($\nu_{ij}=\nu_i+\nu_j$, etc.)
\begin{eqnarray}
 &&
\hspace{-7mm}
I_{2} \propto \sum_{{{m}},{{n}}=0}^{\infty}
 \frac{
(-2-\varepsilon/2+\nu_{12})_m      
(-2-\varepsilon/2+\nu_{45})_n}{(-4-\varepsilon+\nu_{12345})_{n+m}}      
\nonumber\\ &&
\hspace{-7mm}
\cdot \frac {(2+\varepsilon/2)_{m+n}
(2+m+\varepsilon/2)_{-\nu_1} 
(2+n+\varepsilon/2)_{-\nu_5}}
{m! n!},      \nonumber\\
 \end{eqnarray}
with $(a)_c \equiv \Gamma(a+c)/\Gamma(a)$ the Pochhammer symbol.
 We checked this representation 
 for various integer values of the $\nu_i$ using {\sf MATAD} and found 
 complete agreement. Note that for all diagrams deriving
 from this topology, the $F_1$--structure occurs due to the diagram's
 topology and mass distribution, and its form is independent of the operator 
 insertion. As in the previous section and in the $2$--loop case, this allows
 to obtain a representation in terms of a multiply nested sum, which 
 is regularized and can be expanded in $\ep$, for each diagram belonging 
 to this class. Consider as an example the scalar diagram shown in Figure
 \ref{TOP3AEX}, with all $\nu_i=1$.

\vspace*{-5mm}

\vspace*{-4mm}
 One obtains the following parameter integral
 \begin{eqnarray}
  &&\hspace{-6mm} I_{3}=C_3\int_0^1dx_i~
      x_3^{\eph-1}(1-x_3)^\eph
x_4^{\eph-1}
\N\\&&\hspace{-6mm}
     (1-x_4)^\eph (x_5(1-x_4)+x_4x_6(1-x_1-x_2)
\N\\&&\hspace{-6mm}
     +x_4x_1x_7+x_4x_2x_5)^{N-1}
     x_1^{-\eph}x_2^{-\eph}\theta(1-x_1-x_2)
\N\\&&\hspace{-6mm}
     (1-x_1-x_2)
\N\\&&\hspace{-6mm}
     \left(1-x_1\frac{x_3-1}{x_3}-x_2\frac{x_4-1}{x_4}\right)^{-2+3\ep/2}~.
     \label{I3ex}
 \end{eqnarray}
 $C_3$ is a trivial pre-factor. The operator insertion contributes as 
 a polynomial linear in each Feynman parameter raised to the power $N$. 
 This term is in a certain sense ``convoluted'' with a structure deriving from 
 the $F_1$ in Eq. (\ref{F1def}). The application of the binomial theorem implies 
 additional sums, until the integrand is given in such a way that 
 Eq. (\ref{F1def}) can be applied~\footnote{At first sight, the last line
 of Eq. (\ref{I3ex}) and the denominator of Eq. (\ref{F1def}) seem to be different.
 After applying analytic continuation relations for $F_1$, their 
 structure becomes the same}. 
 Thus one obtains:
 \begin{eqnarray}
  &&\hspace{-6mm}I_{3}=       \frac{C_3}{(N+1)(N+2)}
             \sum_{m=0}^{\infty}
             \sum_{n=0}^{\infty}
             \sum_{l=2}^{N+2}    {N+2 \choose l} 
             \sum_{j=2}^{l}      
  \N\\ &&\hspace{-6mm}
             {l \choose j} \Biggl\{
             \sum_{k=1}^{j}    {j \choose k}
             \sum_{r=0}^{l-k}   {l-k \choose r} (-1)^{l+j+k+r}
  \N\\ && \hspace{-6mm}
              B\left(k, m+1+ \frac{\ep}{2}\right) 
              \frac{\Gamma(k+r+m+n+ \frac{\ep}{2})} 
                   {\Gamma(m+1)\Gamma(n+1)\Gamma(k+r+\frac{\ep}{2})}
  \N\\ && \hspace{-6mm}
             \frac{B\left(r+l-1, n+1 +\frac{\ep}{2}\right)}{(N+3-j)}
  \N\\ && \hspace{-6mm}
             \frac{B\left(k+m-\frac{\ep}{2}, r+1+n - \frac{\ep}{2}\right)}
                  {(k+r+1+m+n-\ep)}
  \N\\&&\hspace{-6mm}
    + \sum_{r=0}^{l-j}    {l-j \choose r}(-1)^{l+j+r}
                   B\left(r+l-1,n+1+\frac{\ep}{2}\right)
  \N\\ &&\hspace{-6mm}
               \frac{\Gamma(j+r+m+n+\frac{\ep}{2})}
                    {\Gamma(m+1)\Gamma(n+1)\Gamma(j+r+\frac{\ep}{2})}
             B\left(j, m+1 + \frac{\ep}{2} \right)
  \N\\ &&\hspace{-6mm}
          \frac{B\left(j+m-\frac{\ep}{2},r+1+n-\frac{\ep}{2}\right)}
               {(j+r+1+m+n-\ep) (N+3-j)}\Biggr\}~.
 \end{eqnarray}
 Here, $B(a,b)$ is the Euler--Beta function.
 After expanding in $\ep$, this sum can be reduced 
 by \SigmaP, cf. \cite{CarstenLL} and references therein. 
 One obtains
 \begin{eqnarray}
  &&\hspace{-6mm}I_3= \frac{C_3}{(N+1) (N+2) (N+3)} \Biggl\{
       \frac{1}{6} {S_1^3}
  \N\\ &&\hspace{-6mm}
       +\frac{N^2+12 N+16}{2 (N+1) (N+2)} {S_1^2}
       +\frac{4 (2 N+3)}{(N+1)^2 (N+2)} {S_1}
  \N\\ &&\hspace{-6mm}
       +\frac{8 (2 N+3)}{(N+1)^3 (N+2)}
       + 2 \Biggl[-2^{N+3} +3 - (-1)^N \Biggr] \zeta_3
  \N\\ &&\hspace{-6mm}
       -(-1)^N {S_{-3}}
       +\Biggl[\frac{3 N^2+40
        N+56}{2 (N+1) (N+2)}
       - \frac{1}{2} {S_1} \Biggr] {S_2}
  \N\\ &&\hspace{-6mm}
       -\frac{3 N+17}{3} {S_3}
       -2 (-1)^N {S_{-2,1}}
       -(N+3) {S_{2,1}}
  \N\\ &&\hspace{-6mm}
       +2^{N+4} {S_{1,2}}\left(\frac{1}{2},1;N\right)
  \N\\ &&\hspace{-6mm}
       +2^{N+3} {S_{1,1,1}}\left(\frac{1}{2},1,1;N\right)\Biggr\}+O(\ep)~.
 \end{eqnarray}
 It is interesting to note, that in this case generalized harmonic sums
 appear in the result for one scalar integral. For the $n_f$--terms, this 
 never happened. There, the generalized sums emerged only when splitting 
 up the complete contribution to one diagram into several pieces, without 
 further reference to certain momentum integrals anymore. This could point 
 towards generalized harmonic sums remaining for individual diagrams or even 
 for the complete physical OME. 
 Another interesting point is that the 
 powers $2^N$ lead to divergences as $N\rightarrow\infty$ and are 
 therefore expected to cancel in the physical expression. 
 Details on how to calculate this class of ladder--type diagrams 
 and results for the most complicated scalar integrals will be given 
 in \cite{BHKS}~.
%
%
%
\section{Conclusions}

The heavy flavor 3--loop Wilson coefficients are needed in consistent analyses
of the DIS World data at NNLO. In case of the structure function $F_2(x,Q^2)$, the 
asymptotic representation applies for scales $Q^2/m^2 \geq 10$.
After a larger number of Mellin moments were computed, now the asymptotic massive Wilson
coefficients are calculated for general values of the Mellin variable $N$. Here, we
studied both the bubble- and ladder topologies and developed the corresponding computational
frame. A first class of contributions has been completed in the case of the quarkonic operator
insertions with all contributions to the color factors $T_F^2 n_f C_{A,F}$. 
\section*{Acknowledgments}
\noindent
This work has been supported in part by SFB-TR/9, the EU TMR network HEPTOOLS, 
Austrian
Science Fund (FWF) grants P20162-N18 and P20347-N18, and the Generalitat Valenciana under Grant No.
PROMETEO/2008/069.
We would like to thank K. Chetyrkin, P. Paule, J.~Smith, M. Steinhauser and
J.~Vermaseren for  useful discussions. 

\begin{footnotesize}

\end{footnotesize}
\end{document}